\documentclass[12pt]{iopart}
\usepackage{graphicx} 
\begin{document}

\def\la{\langle}
\def\ra{\rangle}
\def\om{\omega}
\def\Om{\Omega}
\def\vep{\varepsilon}
\def\wh{\widehat}
\def\P0{\wh{\cal P}_0}
\def\dt{\delta t}
\def\wh{\widehat}
\def\da{^\dagger}
\def\nn{\nonumber}
\def\iz{\left}
\def\zi{\right}
\def\vep{\varepsilon}
\def\tx{\tilde{x}}
\def\tik{\tilde{k}}
\def\tk0{\tilde{k}_0}
\newcommand {\bR} {\mathbb{R}}
\newcommand {\cO} {\begin{cal} O \end{cal}}

\newcommand {\fexp} [1] {\exp \left( #1 \right)}
\newcommand {\fabs}[1] {\left| #1 \right|}
\newcommand {\fabsq}[1] {\left| #1 \right|^2}

\newcommand{\beq}{\begin{equation}}
\newcommand{\eeq}{\end{equation}}
\newcommand{\beqa}{\begin{eqnarray}}
\newcommand{\eeqa}{\end{eqnarray}}
\newcommand{\intf}{\int_{-\infty}^\infty}
\newcommand{\into}{\int_0^\infty}
\letter{Physical realization of ${\cal{PT}}$-symmetric potential
scattering in a planar slab waveguide}
\author{A. Ruschhaupt, F. Delgado and J. G. Muga}
\address{Departamento de Qu\'\i{}mica-F\'\i{}sica, Universidad del
Pa\'\i{}s Vasco, Apdo. 644, 48080 Bilbao, Spain}

\begin{abstract}
A physical realization of scattering by ${\cal{PT}}$-symmetric
potentials is provided: 
we show that the Maxwell equations for an electromagnetic wave
travelling along a planar slab waveguide filled with 
gain and absorbing media in contiguous regions, can be approximated in a
parameter range by a Schr\"odinger equation with a ${\cal{PT}}$-symmetric 
scattering potential. 
\end{abstract}
\nosections


${\cal{PT}}$-symmetric Hamiltonians remain invariant under
the combination of 
parity and time reversal symmetry operations. They  
have attracted considerable attention in diverse areas  
such as quantum field theory 
\cite{BM97}, solid state physics \cite{Hollowood92,HN96}, or  
population biology \cite{NS98}. One of their most important
and striking properties is that 
the discrete eigenvalues are real if the eigenstates are also 
${\cal{PT}}$-invariant, or appear in conjugate pairs otherwise. Most  
of the work on ${\cal{PT}}$-invariance has dealt with discrete 
Hamiltonians and much less
attention has been paid to scattering systems at a general or fundamental
level.  
A few studies of the scattering by these 
potentials refer to specific models \cite{LCV01,AFK02},
or examine transparent ${\cal{PT}}$-symmetric potentials \cite{ACDI99,BR00}.
Some generic results, restricted to real momentum and local 
potentials, have been discussed by Deb, Khare and Roy
\cite{DKR03}, whereas a general formal scattering
theory for one dimensional ${\cal{PT}}$-symmetric
potentials is provided in a recent
review about complex 
potentials \cite{review}.


As in \cite{review}, we shall assume that the Hamiltonian of the 
non relativistic particle of mass $m$
can be written as the sum of the kinetic energy operator 
corresponding to the ``free-motion''
evolution, $H_0$, and the  
potential operator $V$,
\beq\label{ham}
H=H_0+V.
\eeq
$V$ may be generically non-local.
Consider the combined action of the anti-unitary time reversal 
operator ${\cal{T}}$ (${\cal{T}} c|x\ra= c^*|x\ra$) and 
the parity unitary operator ${\cal{P}}$
(${\cal{P}} c|x\ra= c|-x\ra$), 
where $c$ is an arbitrary complex constant. 
``${\cal{PT}}$-invariant'' Hamiltonians \cite{BB98} 
remain unchanged by this transformation, 
\beq
[{\cal{P}}{\cal{T}}, H]=0\,.
\eeq
Since the kinetic energy operator $H_0$ is ${\cal{PT}}$-invariant,
this implies
(note the erratum in  \cite{review})
\beq
\la x|V|x'\ra=\la -x'|V^\dagger|-x\ra=\la -x|V|-x'\ra^*\,.
\eeq
In the particular case of local interactions,
i.e., when $\la x|V|x'\ra=\delta(x-x')V(x)$, 
${\cal{PT}}$-symmetry implies $V(x)=V(-x)^*$
and $V^\dagger={\cal{P}} V {\cal{P}}$; the real part of $V(x)$ must be  
symmetric and the imaginary part antisymmetric.
The consequences of ${\cal{PT}}$-symmetry
in the scattering amplitudes (their structure,  pole   
configuration, and relations in the momentum complex plane)
have been examined in \cite{review}.


One important open question is the possible physical meaning of  
non trivial ${\cal{PT}}$-symmetric local interactions, i.e.,  
non-hermitian ones 
with a non-vanishing imaginary part. So far, the existing physical
realizations of ${\cal{PT}}$-symmetry involve
some non-locality, e.g., velocity dependent 
potentials \cite{HN96, NS98}. 
In this letter we show that  
the scattering of electromagnetic waves 
provides a physical realization of a local ${\cal{PT}}$-symmetric
interaction 
since absorption and gain processes
may be implemented at different spatial regions with appropriately 
chosen media. 
%

\begin{figure}
\begin{center}
\includegraphics[width=0.3\linewidth]{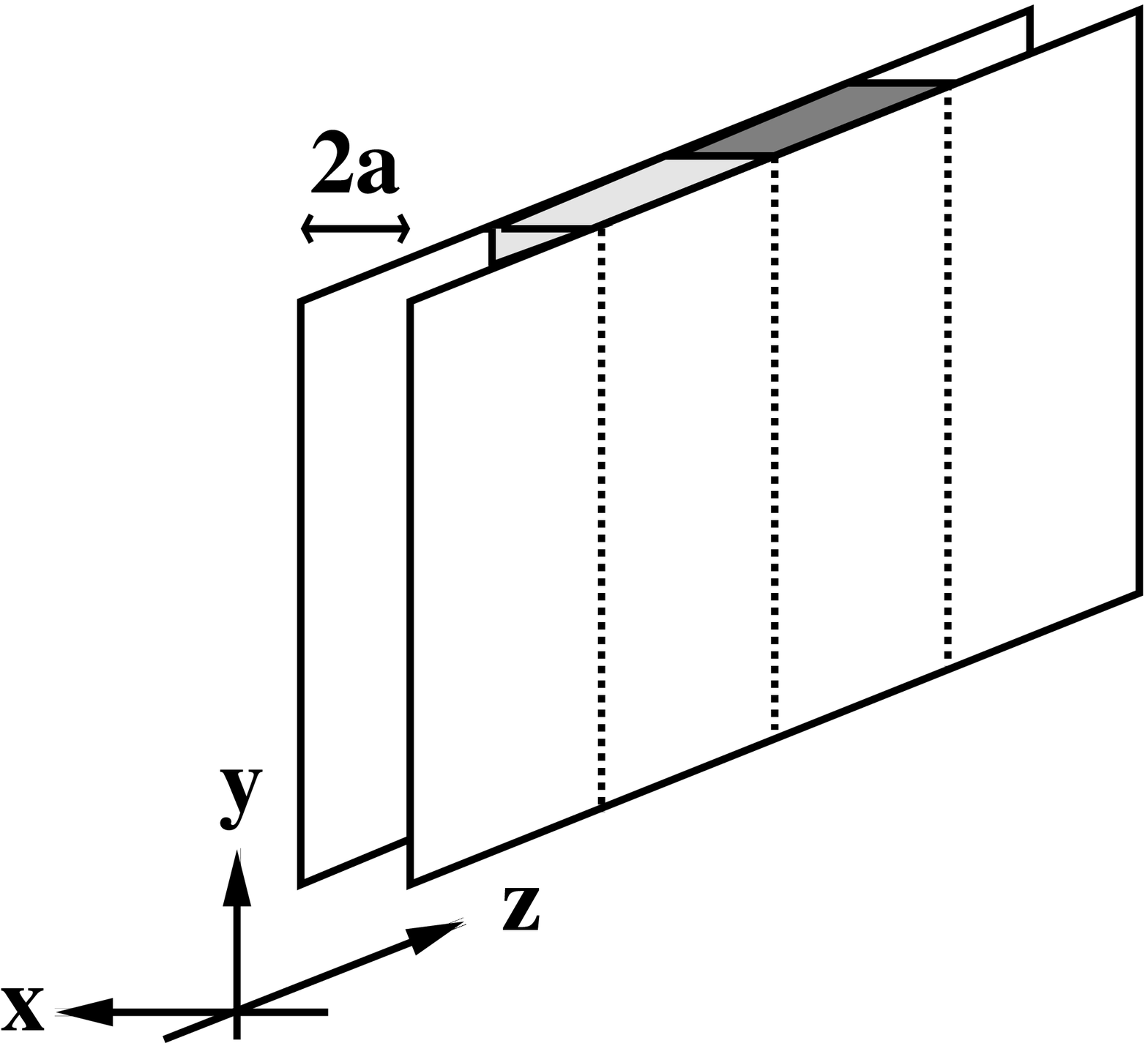}\qquad
\includegraphics[width=0.49\linewidth]{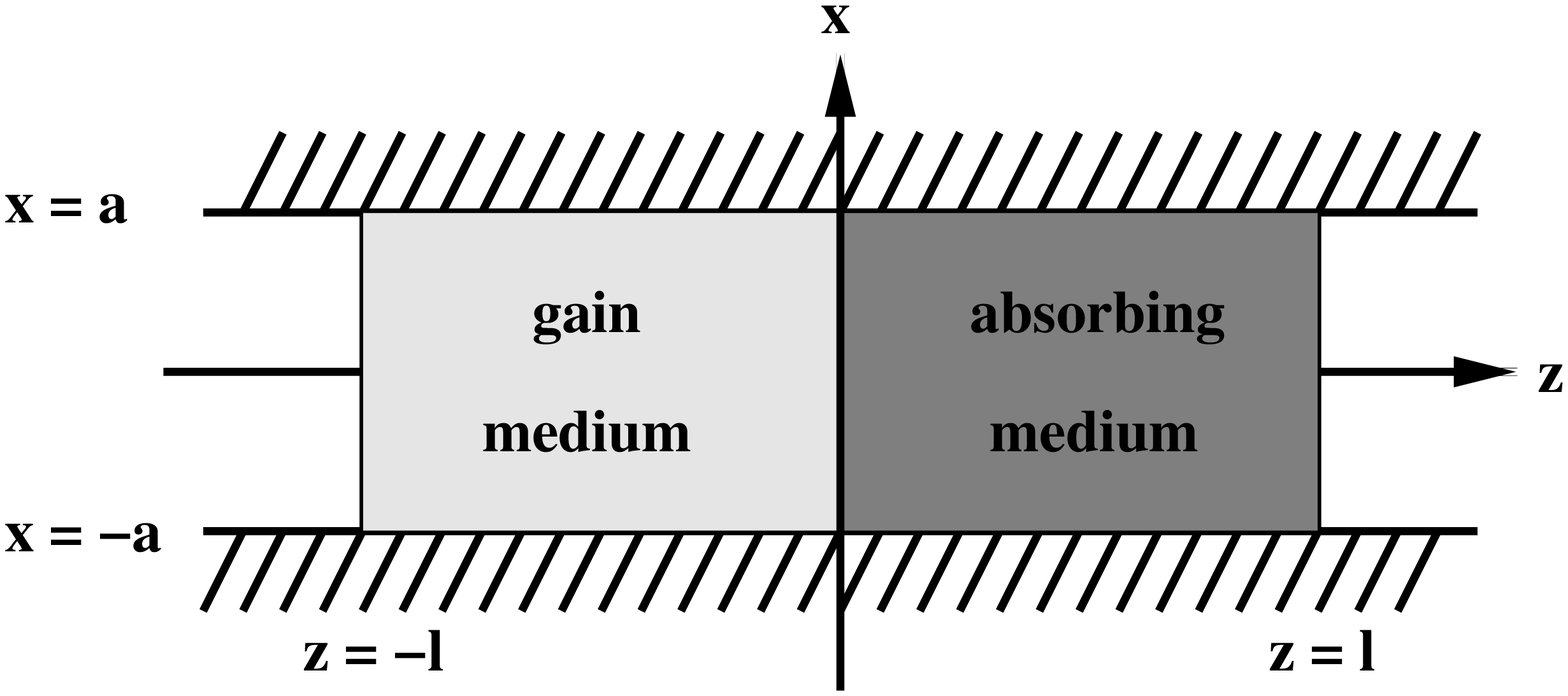}
\end{center}
\caption{\label{fig1}Schemes of the planar slab waveguide: 
three dimensional (left), and in the $xz$-plane (right).}
\end{figure}

Let us consider two  ideal metallic slabs
for $x<-a$ and $x> a$ forming a ``planar waveguide'' in $z$ and $y$
as depicted in Figure 1. In a realistic setting 
the waveguide must of course have also a finite width along
the $y$-direction 
but for a sufficiently large width the  
results of the finite width case may be arbitrarily close to  
those for the infinite one.
In the following we shall only consider
the space between the metallic slabs, $|x|\le a$. 
The waveguide is empty for $|z|>l$, with
relative permittivity $\epsilon=1$; there is also a gain 
medium region $-l < z < 0$ and an absorbing region  
$0 < z < l$. Both regions are filled with the same atomic gas,  
but in the gain region  
a laser shining in $y$ direction
pumps the resonant atoms to their excited state 
to produce  
a population inversion \cite{Chiao93}.
The relative
permittivity, assuming a simple Lorentz model,
can thus be written as  
\beq
\epsilon(z,\omega)
=1-\xi(z)\frac{\omega_p^2}{\omega^2-\omega_0^2+2i\delta\omega}
\eeq
where 
\beq
\xi(z)=\left\{
\begin{array}{rlc}
-1& {\rm{gain\; medium}} & -l < z < 0
\\
1& {\rm{absorbing\; medium}} & \;0 < z < l
\\
0&{\rm vacuum} & |z|>l
\end{array}\right.,
\eeq
$\omega_p$ is the plasma frequency, $\omega_0$ the resonance frequency, 
and $\delta$ a damping constant.
 
The Maxwell equations are
\begin{eqnarray}
\begin{array}{rclcrcl}
\nabla \times \vec{E} + \frac{\partial}{\partial t} \vec{B} & = & 0 & \qquad &
\nabla \times \vec{B} - \frac{1}{c^2} \frac{\partial}{\partial t}
\epsilon \left(z, i\frac{\partial}{\partial t}\right) \vec{E} & = & 0\\
\nabla \cdot \vec{E} & = & 0 & \qquad &
\nabla \cdot \vec{B} & = & 0
\end{array}
\label{maxwell}
\end{eqnarray}
$c$ being the speed of light in vacuum.
We assume now the following form of an electromagnetic wave traveling
in the $z$ direction
\begin{eqnarray}
\begin{array}{rcl}
\vec{E}(x,y,z,t) &=& \int d\omega\, \left(\begin{array}{c}
0 \\ -i\omega \hat{B}_x (x) \phi (z,\omega)\\ 0
\end{array}\right)
e^{-i\omega t}\\
\vec{B}(x,y,z,t) &=& \int d\omega\, \left(\begin{array}{c}
\hat{B}_x (x) \frac{\partial \phi}{\partial z} (z,\omega)\\ 0 \\
-\frac{\partial\hat{B}_x}{\partial x} (x) \phi(z,\omega)
\end{array}\right)
e^{-i\omega t}
\end{array}
\label{ansatz}
\end{eqnarray}
where the physical fields are given by the real part of $\vec{E}$ and
$\vec{B}$.
Inserting the ansatz (\ref{ansatz}) into the Maxwell equations (\ref{maxwell})
we get by separation of variables
\begin{eqnarray}
\left(c^2 \frac{\partial^2}{\partial x^2} + \omega_c^2\right) \hat{B}_x (x)
= 0
\label{eq2}
\end{eqnarray}
and
\begin{eqnarray}
\left(k^2 (z,\omega)
+ \frac{\partial^2}{\partial z^2}\right) \phi (z,\omega) = 0,
\label{eq_exact}
\end{eqnarray}
$\omega_c$ being the cut-off frequency and
$k^2 = \frac{1}{c^2}(\omega^2 \epsilon (z,\omega) - \omega_c^2)$.
The boundary and matching conditions following from classical
electrodynamics \cite{jackson.book} are
$\hat{B}_x (-a) = \hat{B}_x (a) = 0$.
A solution of (\ref{eq2}) fulfilling the boundary conditions is
$\hat{B}_x (x) = \cos(\pi x/2a)$ which results
in $\omega_c = c\pi/2a$.
Moreover it follows from classical
electrodynamics that $\phi (z,\omega)$,
$\frac{\partial\phi}{\partial z} (z,\omega)$
must be continuous for all $z$ and all $\omega$.
From the fact that $\phi (z,\omega)$ is
a solution of (\ref{eq_exact}) it follows that $\frac{\partial^2 \phi}{\partial z^2}
(z,\omega)$ is continuous for all $\omega$ and all $z \not\in \{-l,0,l\}$
($k^2 (z,\omega)$ is non-continuous at $z=-l,0,l$).
We choose the waveguide width $2a$ so that 
the cut-off frequency  
coincides with the resonance frequency, $\omega_c = \omega_0$, 
and 
consider waves with frequencies near the cut-off, 
$\omega=\omega_c+\Delta \omega$.
%
It follows that 
\begin{eqnarray}
\fl k^2 (z, \omega_c + \Delta\omega) &=&
2 \frac{\Delta\omega \, \omega_c}{c^2} (1 + \Delta\omega/2\omega_c)
- \xi(z) \frac{\omega_p^2}{c^2}
\frac{1 + 2\frac{\Delta\omega}{\omega_c}(1+\Delta\omega/2\omega_c)}
{2\frac{\Delta\omega}{\omega_c}(1+\Delta\omega/2\omega_c)
+2i(1+\Delta\omega/\omega_c)\delta/\omega_c}\nonumber\\
&=& i \xi(z) \frac{\omega_c \omega_p^2}{2 c^2 \delta}
+ \frac{\Delta\omega}{\omega_c} \; \frac{2\omega_c^2}{c^2} \left(1 - \xi(z)
\frac{\omega_p^2}{4 \delta^2}
+ i \xi(z) \frac{\omega_p^2}{4 \delta \omega_c}\right) 
+ \cO\left(\frac{\Delta\omega}{\omega_c}\right)^2
\end{eqnarray}
and so, from (\ref{eq_exact}),
\begin{eqnarray}
\fl \frac{\partial^2}{\partial z^2} \phi(z,\omega_c+\Delta\omega)
= - k^2 (z,\omega_c+\Delta\omega)  \phi(z,\omega_c+\Delta\omega)\nonumber\\
\fl = -\left(i \xi(z) \frac{\omega_c \omega_p^2}{2 c^2 \delta}
+ 2 \Delta\omega \frac{\omega_c}{c^2} \left(1 - \xi(z)
\frac{\omega_p^2}{4 \delta^2}
+ i \xi(z) \frac{\omega_p^2}{4 \delta \omega_c}\right) 
+ \cO\left(\frac{\Delta\omega}{\omega_c}\right)^2\right)
 \phi(z,\omega_c+\Delta\omega)
\end{eqnarray}
We assume that $\omega_p^2/\delta \ll \delta, \omega_c$.
For $\Delta \omega \ll \delta,\omega_c$ we get
\begin{eqnarray*}
k^2 (z,\omega_c+\Delta\omega)
\approx \frac{2\omega_c}{c^2} \Delta\omega
+ i \xi(z) \frac{\omega_c\omega_p^2}{2c^2\delta}
=: \tilde{k}^2_{\omega_c} (z, \Delta\omega)
\end{eqnarray*}
and
\begin{eqnarray}
\fl \frac{\partial^2}{\partial z^2} \phi(z,\omega_c+\Delta\omega)
&\approx& -\left(i \xi(z) \frac{\omega_c \omega_p^2}{2 c^2 \delta}
+ 2 \Delta\omega \frac{\omega_c}{c^2} \right)
 \phi(z,\omega_c+\Delta\omega)\nonumber\\
\fl \Rightarrow \Delta\omega\,
\phi(z,\omega_c+\Delta\omega)
 &\approx& -\frac{c^2}{2\omega_c}
\frac{\partial^2}{\partial z^2} \phi(z,\omega_c+\Delta\omega)
- i  \xi(z) \frac{\omega_p^2}{4 \delta}
\phi(z,\omega_c+\Delta\omega).
\label{eq}
\end{eqnarray}
Let $\psi(z,t) := \int d\Delta\omega\; \phi(z,\omega_c+\Delta\omega)
e^{-i\Delta\omega t}$ and we get 
\begin{eqnarray*}
\frac{\partial \psi}{\partial t} (z,t) &=&
\int d\Delta\omega\; \Delta\omega \; \phi(z,\omega_c+\Delta\omega)
e^{-i\Delta\omega t} \qquad \forall z,t\\
\frac{\partial^2\psi}{\partial z^2}  (z,t) &=&
\int d\Delta\omega\; \frac{\partial^2\phi}{\partial z^2} (z,\omega_c+\Delta\omega)
e^{-i\Delta\omega t}\qquad \forall z,t \, \mbox{with} \, z \not\in \{-l,0,l\}
\end{eqnarray*}
Note that $\psi (z,t) $ is related directly to the magnetic field component 
${B}_z (x,y,z,t) = -\frac{\partial \hat{B}_x}{\partial x}(x)
e^{-i\omega_c t} \psi (z,t)$.

We assume that $\phi (z,\omega)=0$ for $\fabs{\omega-\omega_c} \ge \Omega$
for some $\Omega \ll \delta,\omega_c$.
One finds, using (\ref{eq}), 
\begin{eqnarray}
\fl i\frac{\partial \psi}{\partial t} (z,t)
&=& \int d\Delta\omega\; (\hbar\Delta\omega) \; \phi(z,\omega_c+\Delta\omega)
e^{-i\Delta\omega t}\nonumber\\
&\stackrel{(\ref{eq})}{\approx}& \int d\Delta\omega\; 
\left(-\frac{c^2}{2\omega_c} \frac{\partial^2}{\partial z^2} \phi (z, \omega_c+\Delta\omega)
- i  \xi(z) \frac{\omega_p^2}{4 \delta}
\phi(z,\omega_c+\Delta\omega)\right)e^{-i\Delta\omega t}\nonumber\\
&=&-\frac{c^2}{2\omega_c}\int d\Delta\omega\; 
\frac{\partial^2}{\partial z^2} \phi (z, \omega_c+\Delta\omega)e^{-i\Delta\omega t}
-\, i  \xi(z) \frac{\omega_p^2}{4 \delta}
\int d\Delta\omega\; \phi (z, \omega_c+\Delta\omega)e^{-i\Delta\omega t}\nonumber\\
&=&-\frac{c^2}{2\omega_c}\frac{\partial^2}{\partial z^2}\psi (z,t)
-\, i  \xi(z) \frac{\omega_p^2}{4 \delta}\psi (z,t)
\label{eq1}
\end{eqnarray}
To make this equation formally equal to a Schr\"odinger equation we multiply
(\ref{eq1}) with $\hbar$ and introduce a auxiliary ``mass'' $m=\hbar\omega_c/c^2$.
Then we can write (\ref{eq1}) in the following form
\begin{eqnarray}
i\hbar\frac{\partial \psi}{\partial t} (z,t)
= -\frac{\hbar^2}{2m}
\frac{\partial^2}{\partial z^2} \psi + V(z) \psi
\label{eq_approx}
\end{eqnarray}
where we have defined an ``effective potential'' $V (z) = -i \xi(z)
\frac{\omega_p^2\hbar}{4\delta}$ which is ${\cal{PT}}$-symmetric.
But there is also a physical motivated way for this mass.
Quantum mechanically we can view the classical electromagnetic field
as a mean field consisting of many photons. Roughly speaking
we are considering here the case that all photons have approximately
the frequency $\omega_c$. A single photon with frequency $\omega_c$ has
the energy $\hbar\omega_c$ and therefore the momentum $p_c=\hbar\omega_c/c$.
A photon has rest mass zero but nevertheless we can view
the momentum as the product of velocity and a velocity dependent ``mass'' $m$.
In the case of a single photon we get $m=p_c/c=\hbar\omega_c/c^2$.

Conversely, let $\psi (z,t)$ be a continuously differentable solution
of (\ref{eq_approx}) and we
define $\phi(z,\omega):=\frac{1}{2\pi}\int dt\, \psi(z,t) e^{i(\omega-\omega_c)t}$.
If $\phi (z,\omega) < \infty$ for all $z,\omega$ and
$\phi (z,\omega)=0$ for $\fabs{\omega-\omega_c} \ge \Omega$
for some $\Omega \ll \delta,\omega_c$ then
\begin{eqnarray}
\fl \frac{\partial^2\phi}{\partial z^2} (z,\omega)
&=&
\frac{1}{2\pi}\int dt\, \frac{\partial^2}{\partial z^2} )\psi(z,t)
 e^{i(\omega-\omega_c)t}\nonumber\\
&=&
\frac{1}{2\pi}\int dt\, \left(-i \frac{2m}{\hbar}\frac{\partial}{\partial t}\psi(z,t)
+ \frac{2m}{\hbar^2} V(z)\psi(z,t)\right) e^{i(\omega-\omega_c)t}\nonumber\\
&\stackrel{(*)}{=}&
\frac{1}{2\pi}\int dt\, \left(i \frac{2m}{\hbar}\psi(z,t)
\frac{\partial}{\partial t}e^{i(\omega-\omega_c)t}
+ \frac{2m}{\hbar^2} V(z)\psi(z,t) e^{i(\omega-\omega_c)t}\right)\nonumber\\
&=&
- \frac{2m}{\hbar}(\omega-\omega_c)\phi (z,\omega)
+ \frac{2m}{\hbar^2} V(z) \phi(z,\omega)\nonumber\\
&=& -\left(\frac{2\omega_c}{c^2}(\omega-\omega_c) + i \frac{\omega_c\omega_p^2}{2c^2\delta}
\xi(z)\right) \phi (z,\omega)\nonumber\\
&=& - \tilde{k}^2_{\omega_c} (z, \omega-\omega_c) \phi(z,\omega)
\approx - k^2 (z,\omega) \phi(z,\omega)
\label{eq3}
\end{eqnarray}
where we have used partial integration at $(*)$. Note that
a necessary condition for $\phi (z,\omega)$ being always finite is
that $\psi (z,\pm\infty)=0$ so there are no ``boundary'' terms in
the partial integration.
Therefore it follows from (\ref{eq3}) that $\phi (z,\omega)$ is approximately
a solution of (\ref{eq_exact}).

We have checked numerically the validity of the approximation 
assuming a resonance frequency in the ultraviolet region, 
see Figure 2. 
$T^l$ ($T^r$) is defined as the transmission amplitude  
and $R^l$ ($R^r$) as the reflection amplitude for incidence from
the left (right) \cite{review} of a harmonic plane wave.
Note that for a real potential $V(z)$, from the unitarity
of the scattering matrix,  
$|T^l|^2 + |R^l|^2 = |T^r|^2 + |R^r|^2 = 1$, but otherwise
these equalities do 
not hold in general. 
Figure 2 shows the sums $|T^{l,r}|^2 + |R^{l,r}|^2$
calculated exactly, with the Maxwell equations, and approximately, 
i.e., with the effective Schr\"odinger equation (\ref{eq_approx}).  
The approximation is very good for
$1\le\frac{\omega}{\omega_c} < 1.02$, and is still in qualitative agreement 
with the exact result above that value.
Note the dominance of 
gain/absorption for low energy scattering from the left/right, 
corresponding to the region found first by the wave.   

Summarizing, we have shown that the equation for an electromagnetic
wave travelling in a planar slab waveguide can be approximated
in a parameter
region by an effective Schr\"odinger equation with a 
scattering, localized, ${\cal{PT}}$-symmetric potential. This provides 
a physical realization of scattering off 
${\cal{PT}}$-symmetric local potentials and opens a way for 
experimental implementation and testing of theoretical results.

\begin{figure}
\begin{center}
\includegraphics[angle=-90,width=0.49\linewidth]{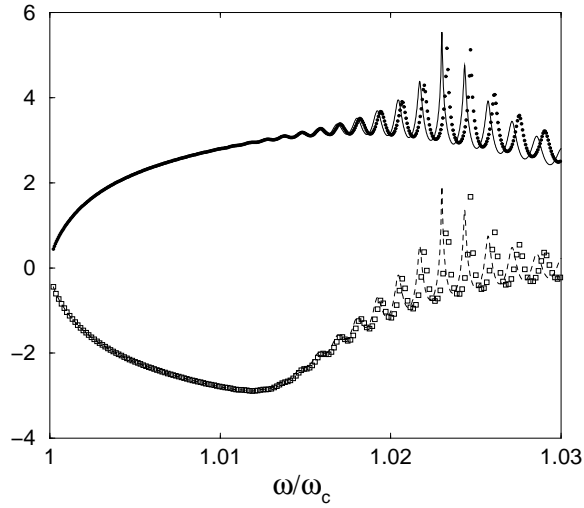}
\end{center}
\caption{\label{fig2}$log_{10}$ of $|T^l|^2 + |R^l|^2$ (solid line) 
and of $|T^r|^2 + |R^r|^2$ (dashed line)
for the exact solution (\ref{eq_exact})
and for the stationary solution of the
approximation (\ref{eq_approx})
(filled and unfilled symbols) versus $\omega/\omega_c$. 
$2a=0.124 \mu m$,
$\hbar\omega_c = \hbar\omega_0 = 5 eV$,
$\hbar\omega_p = 0.2 eV$, $2\hbar\delta = 2.5 eV$, $l=19.7 \mu m$.}
\end{figure}

\ack{We acknowledge A. Steinberg for useful comments. 
This work has been supported
by Ministerio de Ciencia y Tecnolog\'\i a, FEDER (BFM2000-0816-C03-03,
BFM2003-01003), 
and UPV-EHU (00039.310-13507/2001).  
AR acknowledges support of the German
Academic Exchange Service (DAAD),
and Ministerio de Educaci\'on y Ciencia.}


\section*{References}


\begin{thebibliography}{10}

\bibitem{BM97} Bender C M and  Milton K A 1997 {\it Phys. Rev. D}
{\bf 55} R3255

\bibitem{Hollowood92} Hollowood T J 1992 {\it Nucl. Phys. B} {\bf 386} 166

\bibitem{HN96} Hatano N and Nelson D R 1996 {\it Phys. Rev. Lett.} {\bf 77} 570

\bibitem{NS98}  Nelson D R and  Shnerb N M 1998 {\it Phys. Rev. E} {\bf 58} 1383. 

\bibitem{LCV01} L\'evai G, Cannata F and Ventura A 2001
{\it J. Phys. A: Math. Gen.} {\bf 34} 839

\bibitem{AFK02} Albeverio S, Fei S M and Kurasov P 2002
{\it Lett. Math. Phys.} 
{\bf 59} 227 

\bibitem{ACDI99} Andrianov A A, Cannata F, Dedonder J P and Ioffe M V 
1999 {\it Int. J. Mod. Phys. A} {\bf 14} 2675

\bibitem{BR00} Bagchi B and Roychoudhury R 2000
{\it J. Phys. A: Math. Gen.} {\bf 33}
L1 

\bibitem{DKR03}  Deb R N, Khare A and Roy B 2003 {\it Phys. Lett. A} {\bf 307}
215 

\bibitem{review} Muga J G, Palao J P, Navarro B and Egusquiza I L
2004 
{\it Phys. Rep.} {\bf 395} 357

\bibitem{Chiao93} Chiao R Y 1993 {\it Phys. Rev. A} {\bf 48} R34

\bibitem{jo73} Joffily S 1973 {\it Nuclear Physics A} {\bf 215}
301

\bibitem{CSW82} Cassing W, Stingl M and Weiguny A 1982
{\it Phys. Rev. C} {\bf 26} 22

\bibitem{BB98} Bender C M and Boettcher S 1998
{\it Phys. Rev. Lett.} {\bf 80} 5243 



\bibitem{jackson.book}
Jackson J D 1998
{\it Classical Electrodynamics}
(New York: Wiley) 




\end{thebibliography}
\end{document}